# The mass lowest limit of a black hole: the hydrodynamic approach to quantum gravity

Piero Chiarelli

*National Council of Research of Italy, Area of Pisa, 56124 Pisa, Moruzzi 1, Italy*

*Interdepartmental Center "E.Piaggio" University of Pisa*
Phone: +39-050-315-2359
Fax: +39-050-315-2166

Email: pchiare@ifc.cnr.it.

Abstract: In this work the quantum gravitational equations are derived by using the quantum hydrodynamic description. The outputs of the work show that the quantum dynamics of the mass distribution inside a black hole can hinder its formation if the mass is smaller than the Planck's one.
The quantum-gravitational equations of motion show that the quantum potential generates a repulsive force that opposes itself to the gravitational collapse. The eigenstates in a central symmetric black hole realize themselves when the repulsive force of the quantum potential becomes equal to the gravitational one. The work shows that, in the case of maximum collapse, the mass of the black hole is concentrated inside a sphere whose radius is two times the Compton length of the black hole. The mass minimum is determined requiring that the gravitational radius is bigger than or at least equal to the radius of the state of maximum collapse.

## 1. Introduction

One of the unsolved problems of the theoretical physics is that of unifying the general relativity with the quantum mechanics. The former theory concerns the gravitation dynamics on large cosmological scale in a fully classical ambit, the latter one concerns, mainly, the atomic or sub-atomic quantum phenomena and the fundamental interactions [1-9].
The wide spread convincement among physicists that the general relativity and the quantum mechanics are incompatible each other derives by the complexity of harmonizing the two models.
Actually, the incongruity between the two approaches comes from another big problem of the modern physics that is to unify the quantum mechanics [2] with the classical one in which the general relativity is built in.
Although the quantum theory of gravity (QG) is needed in order to achieve a complete physical description of world, difficulties arise when one attempts to introduce the usual prescriptions of quantum field theories into the force of gravity [3]. The problem comes from the fact that the resulting theory is not renormalizable and therefore cannot be utilized to obtain meaningful physical predictions.
As a result, more deep approaches have been proposed to solve the problem of QG such as the string theory the loop quantum gravity [10] and the theory of casual fermion system [11].
Strictly speaking, the QG aims only to describe the quantum behavior of the gravitation and does not mean the unification of the fundamental interactions into a single mathematical framework. Nevertheless, the extension of the theory to the fundamental forces would be a direct consequence once the quantum mechanics and the classical general relativity were made compatible.
The objective of this work is to derive the quantum gravitational equation by using the quantum hydrodynamic approach and give a physical result.
The quantum hydrodynamic formulation describes, with the help of a self-interacting potential (named quantum potential) [12-13] the evolution of the wave function of a particle through two real variables, the spatial particle density $|\Psi|^2$ and its action $S$ that gives rise to the momentum field of the particle $\frac{\partial S}{\partial q^{\tilde{}}} = -p_{\tilde{}} = -(\frac{E}{c}, -p_i)$. The biunique relation between the solution of the standard quantum mechanics and that one of the hydrodynamic model is completed by the quantization that is given by imposing the irrotational condition to the momentum field $p_{\tilde{}}$ [12].

The quantum properties, stemming from the quantum potential, break the scale invariance of the space. This leads to the fact that the laws of physics depend by the size of the problem so that the classical behavior cannot be maintained at a very small scale [12-17] (see appendix A). The aversion of quantum mechanics to the concentration of a particle in a point is due, in the quantum hydrodynamic description, to the so called quantum potential that leads to a larger repulsive force higher is the concentration of the wave packet. If this quantum effect is considered for the BH collapse, it follows that it stops at a certain point. For the collapse of a very small mass this final point will not be beyond the horizon of the events and it will not generate a BH.

Similarly to the classical mechanics, the quantum hydrodynamic equations of motion can be derived by a Lagrangian function, that obeys to the principle of minimum action, and that can be expressed as a function of the energy-impulse tensor.

Thanks to this analogy, the derivation of the gravity equation for a spatial particle mass density that obeys to the quantum law of motion can be straightforwardly obtained .

The paper is organized as follows: in the first section the Lagrangian formulation of the quantum hydrodynamic model in the non-euclidean space is derived. In the second one, the energy-impulse tensor density of the quantum particle mass distribution is formulated for the gravitational equation.

In the last section the smallest mass value of a Schwarzchild BH is calculated.

## 2. The quantum hydrodynamic equations of motion in non-euclidean space

In the first part of this section we will introduce the quantum hydrodynamic equations (QHEs) where, given the wave function $Œ = |Œ| exp[\frac{iS}{\hbar}]$, the quantum dynamics are solved as a function of $|Œ|$ and $S$, where $|Œ|^2$ is the particle spatial density and $\frac{\partial S}{\partial q^{\tilde{}}} = -p_{\tilde{}} = -(\frac{E}{c}, -p_i)$ its momentum.

For the purpose of this work we derive the QHEs by using the Lagrangian approach. This will allow to obtain the impulse-energy tensor for the quantum gravitational equation in a straightforward manner.

The quantum hydrodynamic equations corresponding to the Klein-Gordon one read [18]

$$g^{\tilde{}\epsilon} \frac{\partial S_{(q,t)}}{\partial q^{\tilde{}}} \frac{\partial S_{(q,t)}}{\partial q^{\epsilon}} + \hbar^2 \frac{\partial_{\tilde{}} \partial^{\tilde{}} |Œ|}{|Œ|} - m^2 c^2 = 0 \qquad (1)$$

$$\frac{\partial}{\partial q_{\tilde{}}} \left( |Œ|^2 \frac{\partial S}{\partial q^{\tilde{}}} \right) = \frac{\partial J_{\tilde{}}}{\partial q_{\tilde{}}} = 0 \qquad (2)$$

where

$$S = \frac{\hbar}{2i} ln[\frac{Œ}{Œ*}] \qquad (3)$$

and where

$$J_{\tilde{}} = \frac{i\hbar}{2m} (Œ* \frac{\partial Œ}{\partial q^{\tilde{}}} - Œ \frac{\partial Œ*}{\partial q^{\tilde{}}}) \qquad (4)$$

is the 4-current.

It is worth noting that equation (1) is the hydrodynamic homologous of the classic Hamilton-Jacobi equation (HJE) and that is coupled to the current conservation equation (2) through the quantum potential. Moreover, being in the hydrodynamic analogy

$$\frac{\partial S}{\partial q^\sim} = -p_\sim = -(\frac{E}{c}, -p_i) \tag{5}$$

it follows that

$$J_\sim = (c\rho, -J_i) = -|\Psi|^2 \frac{p_\sim}{m} = \rho \dot{q}_\sim \tag{6}$$

where

$$\rho = -\frac{|\Psi|^2}{mc^2}\frac{\partial S}{\partial t} \tag{7}$$

and where

$$p_\sim = \frac{E}{c^2}\dot{q}_\sim, \tag{8}$$

Moreover, by using (5), equation (1) leads to

$$\frac{\partial S}{\partial q^\sim}\frac{\partial S}{\partial q_\sim} = p_\sim p^\sim = \left(\frac{E^2}{c^2} - p^2\right) = m^2c^2\left(1 - \frac{V_{qu}}{mc^2}\right)$$
$$= m^2\chi^2 c^2\left(1 - \frac{V_{qu}}{mc^2}\right) - m^2\chi^2 \dot{q}^2\left(1 - \frac{V_{qu}}{mc^2}\right) \tag{9}$$

(where $\chi = 1/\sqrt{1-\frac{\dot{q}^2}{c^2}}$) from where it follows that

$$E = \pm m\chi c^2\sqrt{1 - \frac{V_{qu}}{mc^2}} = \sqrt{m^2c^4\left(1 - \frac{V_{qu}}{mc^2}\right) + p^2c^2} \tag{10}$$

(where the minus sign considers the negative energy states (i.e., antiparticles)) where the quantum potential reads

$$V_{qu} = \frac{\hbar^2}{m}\frac{\partial_\sim \partial^\sim |\Psi|}{|\Psi|} \tag{11}$$

and, finally, by using (8) that

$$p_\sim = \pm m\chi \dot{q}_\sim\sqrt{1 - \frac{V_{qu}}{mc^2}} \tag{12}$$

Thence, the quantum hydrodynamic Lagrangian equations of motion read

$$p_\sim = -\frac{\partial L}{\partial \dot{q}^\sim}, \tag{13}$$

$$\dot{p}_{\sim} = -\frac{\partial L}{\partial q^{\sim}} \qquad (14)$$

where

$$L = \frac{dS}{dt} = \frac{\partial S}{\partial t} + \frac{\partial S}{\partial q_i}\dot{q}_i = -p_{\sim}\dot{q}^{\sim} = (\pm)-\frac{mc^2}{\chi}\sqrt{1-\frac{V_{qu}}{mc^2}} \qquad (15)$$

where the lower minus sign still accounts for the antiparticles.

The motion equation can be obtained by inserting $p_{\sim(\dot{q},q)}$ from (13) into (14). The so obtained equation is coupled to the conservation equation (2) through the quantum potential $V_{qu}$.

For $\hbar \to 0$ it follows that $V_{qu} \to 0$ and the classical equations of motion are recovered.

Thence, the hydrodynamic motion equation deriving by (1) (just for matter or antimatter without mixed superposition of states) read

$$\frac{dp_{\sim}}{ds} = \pm\frac{d}{ds}\left(mcu_{\sim}\left(\sqrt{1-\frac{V_{qu}}{mc^2}}\right)\right) = -\frac{\chi}{c}\frac{\partial L}{\partial q^{\sim}}$$
$$= \pm mc\frac{\partial}{\partial q^{\sim}}\sqrt{1-\frac{V_{qu}}{mc^2}} \qquad (16)$$

that leads to

$$\pm mc\sqrt{1-\frac{V_{qu}}{mc^2}}\frac{du_{\sim}}{ds} = \pm\left(-mcu_{\sim}\frac{d}{ds}\left(\sqrt{1-\frac{V_{qu}}{mc^2}}\right) + mc\frac{\partial}{\partial q^{\sim}}\left(\sqrt{1-\frac{V_{qu}}{mc^2}}\right)\right) = \frac{\chi}{c}\frac{\partial T_{\sim}^{\epsilon}}{\partial q^{\epsilon}}, \quad (17)$$

where $ds = \frac{c}{\chi}dt$ and where the quantum energy-impulse tensor $T_{\sim}^{\epsilon}$ reads

$$T_{\sim}^{\epsilon} = (\pm)-\frac{mc^2}{\chi}\sqrt{1-\frac{V_{qu}}{mc^2}}\left(u_{\sim}u^{\epsilon} - u_{\sim}^{\epsilon}\right). \qquad (18)$$

so that, finally, the motion equation reads

$$\frac{du_{\sim}}{ds} = -u_{\sim}\frac{d}{ds}\left(\ln\sqrt{1-\frac{V_{qu}}{mc^2}}\right) + \frac{\partial}{\partial q^{\sim}}\left(\ln\sqrt{1-\frac{V_{qu}}{mc^2}}\right) \qquad (19)$$

where $u_{\sim} = \frac{\chi}{c}\dot{q}_{\sim}$.

It must be noted that the hydrodynamic solutions given by (19) represent an ensemble wider than that of the standard quantum mechanics since not all the field solutions $p_{\sim}$ warrant the existence of the action integral $S$ so that the irrotational condition of the action gradient [12] (similar to the Bohr-Sommerfeld quantization) has to be imposed in order to find the genuine quantum solutions (see appendix B).

Equation (16) (following the method described in appendix B) can be used to find the eigenstates of matter $Œ_{+n}$, by considering the upper positive sign, and of antimatter $Œ_{-n}$, by using the lower minus sign, that allow to obtain the generic wave function $Œ = Œ_+ + Œ_- = \sum_n (a_{+n} Œ_{+n} + a_{-n} Œ_{-n})$, where $Œ_+ = \sum_n a_{+n} Œ_{+n}$ and $Œ_- = \sum_n a_{-n} Œ_{-n}$.

It must be noted that the equations (13-14) describe the quantum evolution of pure matter or antimatter states (as we need for the calculation in section 3.3). The more general treatment including the superposition of states of matter and antimatter is given elsewhere [19].

Finally, for the solution of the gravitational problem, equation (19) in non-euclidean space reads

$$\frac{du_\sim}{ds} - \frac{1}{2} \frac{\partial g_{\}|}}{\partial q^\sim} u^\} u^| = -u_\sim \frac{d}{ds}\left( ln \sqrt{1 - \frac{V_{qu}}{mc^2}} \right) + \frac{\partial}{\partial q^\sim}\left( ln \sqrt{1 - \frac{V_{qu}}{mc^2}} \right) \quad (20)$$

with the conservation equation

$$\frac{1}{\sqrt{-g}} \frac{\partial}{\partial q^\sim} \sqrt{-g} \left( g^{\sim\epsilon} /Œ/ \frac{\partial S}{\partial q^\epsilon} \right) = 0 \quad (21)$$

where

$$V_{qu} = \frac{\hbar^2}{m} \frac{1}{/Œ/\sqrt{-g}} \partial^\sim \sqrt{-g} \left( g^{\sim\epsilon} \partial_\epsilon /Œ/ \right), \quad (22)$$

where $g_{\epsilon\sim}$ is the metric tensor and where $\frac{1}{g} = /g_{\epsilon\sim}/ = -J^2$, where $J$ is the jacobian of the transformation of the Galilean co-ordinates to non-euclidean ones.

## 3.1 The quantum energy-impulse tensor density

Given the hydrodynamic Lagrangian function $\tilde{L} = \int /Œ/^2 \, L \, dV = \int \boldsymbol{L} \, dV$, its spatial density $\boldsymbol{L}$ reads

$$\boldsymbol{L} = \frac{u\tilde{L}}{uV} = /Œ/^2 \, L \quad (23)$$

that, by using the variational calculus, leads to the quantum impulse energy tensor density (QEITD) [16]

$$
\begin{aligned}
T_{\sim}^{\,\epsilon} &= \dot{q}_{\sim}\frac{\partial L}{\partial \dot{q}_{\epsilon}} - L\mathsf{u}_{\sim}^{\,\epsilon} = |\mathrm{Œ}|^2\left(\dot{q}_{\sim}\frac{\partial L}{\partial \dot{q}_{\epsilon}} - L\mathsf{u}_{\sim}^{\,\epsilon}\right) \\
&= |\mathrm{Œ}|^2\left(-\dot{q}_{\sim}p^{\epsilon} - L\mathsf{u}_{\sim}^{\,\epsilon}\right) = |\mathrm{Œ}|^2\left(\mp\frac{cu_{\sim}}{\chi}mcu^{\epsilon}\sqrt{1-\frac{V_{qu}}{mc^2}} \pm \frac{mc^2}{\chi}\sqrt{1-\frac{V_{qu}}{mc^2}}\mathsf{u}_{\sim}^{\,\epsilon}\right) \\
&= \mp\frac{mc^2|\mathrm{Œ}|^2}{\chi}\sqrt{1-\frac{V_{qu}}{mc^2}}\left(\frac{c}{\chi}u_{\sim}u^{\epsilon}-\mathsf{u}_{\sim}^{\,\epsilon}\right) \\
&= |\mathrm{Œ}|^2\left(\mp\frac{cu_{\sim}}{\chi}mcu^{\epsilon}\sqrt{1-\frac{V_{qu}}{mc^2}} \pm \frac{mc^2}{\chi}\sqrt{1-\frac{V_{qu}}{mc^2}}\mathsf{u}_{\sim}^{\,\epsilon}\right) = |\mathrm{Œ}|^2\,\mathsf{T}_{\sim}^{\,\epsilon}
\end{aligned}
\tag{24}
$$

that reads

$$
\mathsf{T}_{\sim}^{\,\epsilon} = \pm\frac{mc^2|\mathrm{Œ}_{\pm}|^2}{\chi}\sqrt{1-\frac{V_{qu}}{mc^2}}\left(u_{\sim}u^{\epsilon}-\mathsf{u}_{\sim}^{\,\epsilon}\right)
\tag{25}
$$

where

$$
m|\mathrm{Œ}_{\pm}|^2
\tag{26}
$$

are the mass densities of matter or antimatter where the minus sign refers to antimatter.

### 3.2 The quantum gravitational equation for spinless uncharged particles

Equation (19) in the classical limit (i.e., $\hbar \to 0, V_{qu} \to 0$) gives

$$
mc\frac{du_{\sim}}{ds} = \frac{dp_{\sim}}{ds} = -\frac{\partial \mathsf{T}_{\sim}^{\,\epsilon}}{\partial q^{\epsilon}}
\tag{27}
$$

with

$$
\lim_{\hbar \to 0}\mathsf{T}_{\sim}^{\,\epsilon} = \pm\frac{mc^2}{\chi}\left(u_{\sim}u^{\epsilon} - \mathsf{u}_{\sim}^{\,\epsilon}\right).
\tag{28}
$$

Moreover, since

$$
\frac{\partial\frac{mc^2}{\chi}\mathsf{u}_{\sim}^{\,\epsilon}}{\partial q^{\epsilon}} = 0\,,
\tag{29}
$$

it follows that the energy-impulse tensor leads to the same mass motion of the classical one that reads

$\mathsf{T}_{\sim}^{\,\epsilon} = \frac{mc^2}{\chi}u_{\sim}u^{\epsilon}$ (given that the PD behaves like dust matter [12]).

Just from the mechanical point of view, thence, the impulse energy tensor has a freedom of choice so that all tensors $\mathsf{T}_{\epsilon}^{\,\sim} \equiv \mathsf{T}_{\epsilon}^{\,\sim} + L_{(\dot{q},t)}\mathsf{u}_{\epsilon}^{\,\sim}$ lead to the same motion of matter (in a space with fixed geometry).

On the other hand, from gravitaional point of view, the curvature of space associated to the QEITDs of type

$$T_\epsilon^{\ \sim} \equiv T_\epsilon^{\ \sim} + \Lambda_{(\dot{q},t)} u_\epsilon^{\ \sim} \qquad (30)$$

would be different as a function of $\Lambda_{(\dot{q},t)}$. Therefore to end with the correct form of $\Lambda_{(\dot{q},t)}$ we must require that the classical Einstein equation as well as the correct Galilean gravitational field must be recovered in the classical limit.

By imposing this condition the explicit expression

$$\Lambda = \frac{8fG}{c^4} \frac{m |Œ|^2 c^2}{\chi} \qquad (31)$$

is obtained.

Thence, quantum gravitational equation for particles and antiparticles respectively reads [20]

$$R_{\epsilon\sim} - \frac{1}{2} g_{\epsilon\sim} R_r^{\ r} = \frac{8fG}{c^4}\left( T_{\epsilon\sim} + \frac{m |Œ_+|^2 c^2}{\chi} g_{\sim\epsilon} \right) \qquad (32)$$

$$R_{\epsilon\sim} - \frac{1}{2} g_{\epsilon\sim} R_r^{\ r} = -\frac{8fG}{c^4}\left( T_{\epsilon\sim} + \frac{m |Œ_-|^2 c^2}{\chi} g_{\sim\epsilon} \right) \qquad (33)$$

where $T_{\epsilon\sim} \equiv T_\epsilon^{\ r} g_{r\sim}$.

In the classical limit, where particles are localized and distinguishable, we can approximate them by the point-like distribution

$$|Œ_+|^2 = \sum_{a_+} u(r - r_{a_+}), \qquad (34)$$

or

$$|Œ_-|^2 = \sum_{a_-} u(r - r_{a_-}), \qquad (35)$$

while in the quantum case they are defined by the solution of the quantum equation.

Moreover, if in the classical gravity, the equation (32) defining the tensor $g_{\epsilon\sim}$, has to be solved with the mass motion equation (19) (given that $g_{\epsilon\sim}$ itself depends by the motion of the masses) in the quantum case the set up is a little bit more complicated since the motion equation (19) as well as the gravitational equations (32-33) are coupled to the mass conservation equations (21) through $|Œ|$ that is present into the quantum potential.

Finally, noting that the quantum motion equation (19) is equivalent to the HJE equation (1) (see appendix C) and that, with the irrotational condition of the action gradient, equations (1,19) lead to the same solutions of the Klein-Gordon equation [18], we can write the equations of quantum gravity in the standard notations as

$$R_{\epsilon\sim} - \frac{1}{2} g_{\epsilon\sim} R_r^{\ r} = \pm\frac{8fG}{c^4}\left( T_{\epsilon\sim} + \frac{m |Œ_\pm|^2 c^2}{\chi} g_{\sim\epsilon} \right) \qquad (36)$$

$$g_{\sim\epsilon} \partial^\epsilon \partial^\sim Œ = -\frac{m^2 c^2}{\hbar^2} Œ \qquad (37)$$

with

$$T_{\sim\epsilon} = \mp \frac{mc^2 /Œ_{\pm}/^2}{\chi}\left[\sqrt{1-\frac{V_{qu}}{mc^2}}g_{\sim\epsilon} + \sqrt{1-\frac{V_{qu}}{mc^2}}^{-1}\left(\frac{\hbar}{2mc}\right)^2 \frac{\partial ln[\frac{Œ}{Œ*}]}{\partial q^{\sim}}\frac{\partial ln[\frac{Œ}{Œ*}]}{\partial q_{\}}}g_{\}\epsilon}\right] \quad (38)$$

### 3.3 Quantum dynamics in a central symmetric gravitational field

In the classical gravity, the dynamics in a central symmetric gravitational field is simplified if the symmetry is maintained along the evolution of the motion. For the quantum case, the condition of central symmetry has to be owned by the eigenfunctions. The same criterion applies to the hydrodynamic motion equations so that the stationary equilibrium condition, that characterizes the eigenstates, has a central symmetric geometry.
Due to the quantum potential form that generates a repulsive force when the matter concentrates itself more and more, the point-like gravitational collapse in the center of such a black hole is not possible in the quantum case.
In order to investigate this aspect, it is useful to note that the quantum gravitational equations, without the quantum potential, perfectly realize the case of motion of incoherent matter [12]. In this case the solution depends by the mass distribution and by the radial velocity. In classical gravity, the solution can be expressed in a synchronous system in quiet with all masses [21] following the identity

$$\frac{Du_{\sim}}{ds} = 0 \quad (39)$$

that is

$$\frac{Du_{\sim}}{ds} = \frac{du_{\sim}}{ds} - \frac{1}{2}\frac{\partial g_{\}|}}{\partial q^{\sim}}u^{\}}u^{|} = 0 \quad (40)$$

so that, for inward radial velocity (i.e., $u_1 < 0$ where $u_{\sim} = (\chi, \dot{r}, 0, 0)$), it follows that

$$\frac{du_{\sim}}{ds} = \frac{1}{2}\frac{\partial g_{\}|}}{\partial q^{\sim}}u^{\}}u^{|} \quad (41)$$

that, considering the last infinitesimal shell of matter that collapses in a central gravitational field, leads to [18]

$$\frac{du_1}{ds} = \frac{1}{2}\frac{\partial g_{00}}{\partial q^1}u^0 u^0 + \frac{1}{2}\frac{\partial g_{11}}{\partial q^1}u^1 u^1 = -\frac{c}{r^2}\chi^2 + \frac{1}{2(r+c)^2}(u_1)^2 \to -\infty \quad (42)$$

with r that approaches to zero leading to a point-like collapse in the center of the BH [21].
In the quantum case we can observe that the dynamics approach the classical output (41) for large masses since it holds $V_{qu} \to \propto \frac{1}{m}$ .
On the other hand, for mass concentration on very short distances when the quantum potential grows in a sensible manner and can be of order of $mc^2$, it can give an appreciable inertial contribution in the motion equation (20) through the term

$$\frac{\partial}{\partial q^{\sim}}\left(ln\sqrt{1-\frac{V_{qu}}{mc^2}}\right), \tag{43}$$

so that the departure from the classical output is expected.
Following the quantum hydrodynamic protocol [12] (see appendix C) the eigenstates are defined by their stationary "equilibrium" condition that reads

$$u_{\sim} = (1,0,0,0) \tag{44}$$

$$\frac{du_{\sim}}{ds} = 0 \tag{45}$$

The condition of null total force (45) is achieved when the quantum force (i.e., minus the gradient of the quantum potential) is equal and contrary to the external ones (see example in appendix C).
In the quantum case, the presence of quantum potential does not allow us to write the Einstein equation in a synchronous system. Therefore, we can only impose the central symmetry that reads [18,21]

$$ds^2 = e^{\epsilon} c^2 dt^2 - r^2\left(d\eta^2 + \sin^2\eta \, d\xi^2\right) - e^{\vartheta} dr^2 \tag{46}$$

where $q_{\sim} = (ct, r, \eta, \xi)$ and

$$g_{00} = e^{\epsilon} \,;\, g_{11} = -e^{\vartheta} \,;\, g_{22} = -r^2 \,;\, g_{33} = -r^2 \sin^2\eta \,;\, \sqrt{-g} = \left|e^{\frac{\vartheta+\epsilon}{2}} r^2 \sin^2\eta\right|^{-1}; \tag{47}$$

that inserted into the gravity equation leads to [21]

$$\frac{8fG}{c^4}\left(T_1^{\ 1} + \frac{m|\Psi_+|^2 c^2}{\chi}\right) = -e^{-\vartheta}\left(\frac{\epsilon'}{r} + \frac{1}{r^2}\right) + \frac{1}{r^2} \tag{48}$$

$$\frac{8fG}{c^4}\left(T_0^{\ 0} + \frac{m|\Psi_+|^2 c^2}{\chi}\right) = -e^{-\vartheta}\left(\frac{1}{r^2} - \frac{\vartheta'}{r}\right) + \frac{1}{r^2} \tag{49}$$

$$\frac{8fG}{c^4}T_0^{\ 1} = -e^{-\vartheta}\frac{\dot{\vartheta}}{r}. \tag{50}$$

where the apex and the dot over the letter mean derivation respect to r and ct, respectively. Moreover, the quantum potential in this case reads

$$V_{qu} = -\frac{\hbar^2}{m}\frac{1}{|\Psi|\sqrt{-g}}\partial^1\sqrt{-g}\left(e^{-\vartheta}\partial_1|\Psi|\right) \tag{51}$$

It is worth noting that for $m \to \infty$ the gravitational radius $R_g = \frac{2Gm}{c^2}$ goes to infinity while the radius $R_0$, representing the sphere inside which the mass concentrate itself in the stationary equilibrium state, goes to zero since $V_{qu} \propto \frac{1}{m} \to 0$. In this case, the point-like collapse up to (macroscopically speaking) $R_0 = 0$ is possible.

On the other hand, when $m \to 0$ the gravitational radius $R_g$ tends to zero, while both the quantum potential $V_{qu} \propto \dfrac{1}{m}$ and, hence, the radius $R_0$ may sensibly grow.

Moreover, given that to have a BH, all the mass has to be contained inside the gravitational radius, it follows that the minimal allowable mass $m_{min}$ for a BH is the smallest one for which it holds the condition $R_0 \leq R_g$ .

Being $R_0(m_{min})$ the highest value of $R_0$ smaller than $R_g$ , thence, for $R_0 < r \cong R_g$ (with $R_0 \to R_g$ )the quantum potential can approximately read (see appendix D)

$$V_{qu} = -\frac{\hbar^2}{m} \frac{1}{/\mathcal{E}/\sqrt{-g}} \partial^1 \sqrt{-g} \left(e^{-\}} \partial_1 /\mathcal{E}/\right) \cong mc^2 \tag{52}$$

Assuming that in the stationary equilibrium distribution (eigenstate) the mass is concentrated in a sphere of radius $R_0$ for $r > R_0$ we can use the gravitational equation with the approximation of null mass that reads [21]

$$-e^{-\}} \left(\frac{\mathcal{E}'}{r} + \frac{1}{r^2}\right) + \frac{1}{r^2} \cong 0 \tag{53}$$

$$-e^{-\}} \left(\frac{1}{r^2} - \frac{\}'}{r} + \right) + \frac{1}{r^2} \cong 0 \tag{54}$$

$$-e^{-\}} \frac{\dot{\}}}{r} \cong 0 \tag{55}$$

$$\} + \mathcal{E} = 0 \tag{56}$$

$$g_{11} = -e^{\}} = -e^{-\mathcal{E}} = -\left(1 - \frac{R_g}{r}\right)^{-1} \tag{57}$$

$$g = -r^4 \sin^2 [ \tag{58}$$

from where, for $r > R_0$ and $r \cong R_g$ , by (52) it follows that

$$\frac{1}{/\mathcal{E}/r^2} \partial^1 r^2 \left(\left(\frac{R_g}{r} - 1\right) \partial_1 /\mathcal{E}/\right) \approx \left(\frac{mc}{\hbar}\right)^2 \tag{59}$$

and hence that

$$\partial^1 \left(r^2 \left(\frac{R_g}{r} - 1\right)\right) >> r^2 \left(\frac{R_g}{r} - 1\right), \tag{60}$$

leading to approximated equation

$$\frac{1}{|Œ|r^2} \partial^1 r^2 \left(\left(\frac{R_g}{r} - 1\right) \partial_1 |Œ|\right) \cong \frac{1}{r^2}\left(\partial^1 r^2 \left(\frac{R_g}{r} - 1\right)\right) \partial_1 \ln |Œ| \cong \left(\frac{mc}{\hbar}\right)^2 \quad R_0 < r \cong R_g. \quad (61)$$

Moreover, by setting $r = R_g + v$ with $v \ll R_g$, (61) reads

$$\partial_1 \ln |Œ| \cong -\left(\frac{mc}{\hbar}\right)^2 r\left(1 + \frac{v}{R_g}\right) \quad (62)$$

leading to the zero-order approximated solution

$$|Œ| \cong |Œ|_0 \, exp[\,-\frac{r^2}{a^2}\,] \quad (63)$$

where

$$a = \frac{\hbar}{mc} \quad (64)$$

equals the Compton length of the BH.
Moreover, since in order to have a BH, all the mass must be inside the gravitational radius, by posing $R_0 \approx 2a$, from (64) it follows that $R_0 = \frac{2\hbar}{mc} < R_g$ leading to the condition

$$\frac{\hbar}{mcR_g} = \frac{\hbar c}{2m^2 G} = \frac{m_p^2}{2m^2} < \frac{1}{2} \quad (65)$$

and, hence, to

$$m > m_p \quad (66)$$

where $m_p = \sqrt{\frac{\hbar c}{G}}$.

## 4. Comments

Even if the hydrodynamic description was formulated contemporaneously to the Schrödinger equation [19], due to the low mathematical manageability, it is much less popular that the latter.
Nevertheless, the interest in the quantum hydrodynamic model has been never interrupted since its formulation by Madelung [22-25]. This because it has proven to be very effective in describing systems larger than a single atom where fluctuations and quantum decoherence become important in defining their evolution [26].
Moreover, due to the classical-like form, the hydrodynamic description is suitable for the connection between quantum concepts (probabilities) and classical ones such as trajectories [27-29].
The property of the hydrodynamic quantum description of being a bridge between the quantum mechanics and the classical one, allows a straightforward generalization of the Einstein gravity (a pure classical theory) to the quantum case, leading to a model with clear mathematical statements.
Furthermore, since the hydrodynamic approach, once the irrotational condition of the action gradient is applied, becomes equivalent to the quantum one [12,25], the results can be expressed in the standard quantum formalism with a set of equations that are independent by the hydrodynamic approach and that appear well defined.
The hydrodynamic quantum gravity has shown to succeed to determine the minimal mass of a black hole.

The model depicts the quantum gravitational behavior in a classical-like way generalizing it with the help of the self-interaction given by the quantum potential.

## 5. Conclusions

In this work the quantum gravitational equations are derived by using the quantum hydrodynamic description. The work shows that, in the case of maximum gravitational compression (when the repulsive force of the quantum potential is equal to the gravitational one) the BH mass is practically concentrated inside a sphere whose radius $R_0 = \frac{2\hbar}{mc}$ is two times the Compton length of the black hole. The minimum BH mass, equal to the Planck mass $m_p = \sqrt{\frac{\hbar c}{G}}$, follows by requiring that the gravitational radius $R_g = \frac{2Gm}{c^2}$ must be bigger than $R_0$.

# Appendix A

## The quantum potential and the breaking of the classical scale invariance of space

In this section we illustrate how the vacuum properties on small scale are affected by the quantum potential. One of the physical quantities that clearly show breaking of scale invariance of vacuum is the spectrum of the vacuum fluctuations.

The quantum potential finds its definition in the frame of the quantum hydrodynamic representation. For sake of simplicity, we analyze here the hydrodynamic motion equations in the low velocity limit. The generalization to the relativistic limit is straightforward since the expression of the quantum potential remains unaltered.

In the quantum hydrodynamic approach, the motion of the particle density $n_{(q,t)} = |\psi|^2_{(q,t)}$, with velocity $\dot{q} = \frac{\nabla S_{(q,t)}}{m}$, is equivalent to the quantum problem (Schrödinger equation) applied to a wave function $\psi_{(q,t)} = |\psi|_{(q,t)} \exp[\frac{i}{\hbar} S_{(q,t)}]$, and is defined by the equations [12]

$$\partial_t n_{(q,t)} + \nabla \bullet (n_{(q,t)} \dot{q}) = 0, \tag{A.1}$$

$$\dot{q} = \frac{\partial H}{\partial p} = \frac{p}{m} = \frac{\nabla S_{(q,t)}}{m}, \tag{A.2}$$

$$\dot{p} = -\nabla (H + V_{qu}), \tag{A.3}$$

$$S = \int_{t_0}^{t} dt \left( \frac{p \bullet p}{2m} - V_{(q)} - V_{qu(n)} \right) \tag{A.4}$$

where the Hamiltonian of the system is $H = \frac{p \bullet p}{2m} + V_{(q)}$ and where $V_{qu}$ is the quantum potential that reads

$$V_{qu} = -\left(\frac{\hbar^2}{2m}\right) n^{-1/2} \nabla \bullet \nabla n^{1/2}. \tag{A.5}$$

For macroscopic objects (when the ratio $\frac{\hbar^2}{2m}$ is very small) the limit of $\hbar \to 0$ can be applied and equations (A.1-A.4) lead to the classical equation of motion. Even, such simplification *tout court* is not mathematically correct, the stochasticity must be introduced to justify it [14,16].

Actually, since the non local characteristics of quantum mechanics can be generated also by an infinitesimal quantum potential, it can be disregarded when random fluctuations overcame it and produce quantum decoherence [14,16,30].

If we consider the fluctuations of the variable $n_{(q,t)} = |\psi|^2_{(q,t)}$ in the vacuum, as shown in ref.[14-16] equation (1) can be derived as the deterministic limit of the stochastic equation

$$\partial_t n_{(q,t)} = -\nabla \cdot (n_{(q,t)} \dot{q}) + y_{(q,t,T)} \tag{A.6}$$

For the sufficiently general case, to be of practical interest, $y_{(q,t,T)}$ can be assumed Gaussian with null correlation time and independent noises on different co-ordinates. In this case, the stochastic partial differential equation (A.6) is supplemented by the relation [16]

$$<y_{(q_r,t)}, y_{(q_s+\lambda, t+\tau)}> = <y_{(q_r)}, y_{(q_s)}> G(\lambda)\mu(\tau)\delta_{rs} \tag{A.7}$$

where $<y_{(q_r)}, y_{(q_s)}> \propto kT$ [16] where $T$ is the amplitude parameter of the noise (e.g., the temperature of an ideal gas thermostat in equilibrium with the vacuum [14,16]) and $G(\lambda)$ is the shape of the spatial correlation function of the noise $y$.

In order that the energy fluctuations of the quantum potential do not diverge, the shape of the spatial correlation function cannot be a delta-function (so that the spectrum of the spatial noise cannot be white) but owns the the correlation function

$$\lim_{T \to 0} G(\lambda) = exp[-(\frac{\lambda}{\lambda_c})^2] \tag{A.8}$$

The noise spatial correlation function (A.8) is a direct consequence of the PD derivatives of the quantum potential that give rise to an elastic-like contribution to the system energy that reads

$$\overline{H}_{qu} = \int_{-\infty}^{\infty} n_{(q,t)} V_{qu(q,t)} dq = -\int_{-\infty}^{\infty} n_{(q,t)}^{1/2} (\frac{\hbar^2}{2m}) \nabla \cdot \nabla n_{(q,t)}^{1/2} dq, \tag{A.9}$$

where large derivatives of $n(q,t)$ generate high quantum potential energy. This can be verified by calculating the quantum potential values due to the sinusoidal fluctuation of the wave function in the vacuum (i.e., $V_{(q)} = 0$) (e.g., mono-dimensional case)

$$\psi = \psi_0 \cos\frac{2\pi}{\lambda} q \tag{A.10}$$

that leads to

$$V_{qu} = -(\frac{\hbar^2}{2m})(\cos^2\frac{2\pi}{\lambda}q)^{-1/2} \nabla \cdot \nabla (\cos^2\frac{2\pi}{\lambda}q)^{1/2} = \frac{\hbar^2}{2m}(\frac{2\pi}{\lambda})^2 \tag{A.11}$$

showing that the energy of the quantum potential grows as the inverse squared of the the wave length of fluctuation.

Therefore, the presence of components with near zero wave length $\lambda$ into the spectrum of fluctuations can lead to fluctuations of quantum potential with finite amplitude even in the case of null noise amplitude (i.e., $T \to 0$).

In this case the deterministic limit (A.1-A.3) contains additional solutions to the standard quantum mechanics (since fluctuations of the quantum potential would not be suppressed).

Thence, from the mathematical inspection of stochastic equation (A.6-A.7) it comes out that in order to obtain the quantum mechanics on microscopic scale, the additional conditions (A.8) must be included to the set of the stochastic equations of the hydrodynamic quantum mechanics [14-16].

A simple derivation of the correlation function (A.8) can come by considering the spectrum of the PD fluctuations of the vacuum. Since each component of spatial frequency $k = \frac{2\pi}{\lambda}$ brings the energy contribution of quantum potential (A.11), the probability that it happens is

$$p = exp\left[-\frac{E}{kT}\right] = exp\left[-\frac{<V_{(q)} + V_{qu}>}{kT}\right] \quad (A.12)$$

that, for the empty vacuum (i.e., $V_{(q)} = 0$), leads to the expression:

$$p \propto exp\left[-\frac{<Vqu>}{kT}\right] = exp\left[-\frac{<\frac{\hbar^2}{2m}\left(\frac{2\pi}{\lambda}\right)^2>}{kT}\right] \quad (A.13)$$

$$= exp\left[-\frac{\hbar^2}{2mkT}\left(\frac{2\pi}{\lambda}\right)^2\right] = exp\left[-\left(\frac{\pi\lambda_c}{\lambda}\right)^2\right] = exp\left[-\frac{\hbar}{2mc}\frac{\hbar c}{\lambda kT}\right]$$

where

$$\lambda_c = 2\frac{\hbar}{(2mkT)^{1/2}} \quad (A.14)$$

From (A.13) it follows that the spatial frequency spectrum $S(k) \propto p(\frac{2\pi}{\lambda})$ of the vacuum fluctuations is not white.

Fluctuations with smaller wave length have larger energy (and lower probability of happening) so that when $\lambda$ is smaller than $\lambda_c$ their amplitude goes quickly to zero.

Given the spatial frequency spectrum $S(k) \propto p(\frac{2\pi}{\lambda})$, the spatial correlation function of the vacuum fluctuation reads

$$G_{(\lambda)} \propto \int_{-\infty}^{+\infty} exp[ik\lambda] S_{(k)} dk \propto \int_{-\infty}^{+\infty} exp[ik\lambda] exp\left[-\left(k\frac{\lambda_c}{2}\right)^2\right] dk$$

$$\propto \frac{\pi^{1/2}}{\lambda_c} exp\left[-\left(\frac{\lambda}{\lambda_c}\right)^2\right] \quad (A.15)$$

that gives (A.8).

The fact that the vacuum fluctuations do not have a white spectrum but have a length "built in" (i.e., the De Broglie thermal wavelength $\lambda_c$) shows the breaking of the its scale invariance: The properties of the space on a small scale are very different from those ones we know on macroscopic scale. When the physical length of a system is smaller than $\}_c$, the deterministic limit of (A.6) (i.e., the quantum mechanics) applies [31] and we have the emerging of the quantum behavior [16].

# Appendix B

## Analysis of the quantization condition in the quantum hydrodynamic description

If we look at the mathematical manageability of QHEs of quantum mechanics (A.1-A.5) no one would consider them.

Nevertheless, the QHEs attract much attention by researchers. The motivation resides in the formal analogy with the classical mechanics that is appropriate to study those phenomena connecting the quantum behavior and the classical one.

In order to establish the hydrodynamic analogy, the gradient of action (A.4) has to be considered as the momentum of the particle. When we do that, we broaden the solutions so that not all solutions of the hydrodynamic equations can be solutions of the Schrödinger problem.

As well described in ref.[12], the state of a particle in the QHEs is defined by the real functions $|Œ|^2 = n_{(q,t)}$ and $p = \nabla S_{(q,t)}$.

The restriction of the solutions of the QHEs to those ones of the standard quantum problem comes from additional conditions that must be imposed in order to obtain the quantization of the action.

The integrability of the action gradient, in order to have the scalar action function $S$, is warranted if the probability fluid is irrotational, that being

$$S_{(q,t)} = \int_{q_0}^{q} dl \cdot \nabla S = \int_{q_0}^{q} dl \cdot p \qquad (B.1)$$

is warranted by the condition

$$\nabla \times p = 0 \qquad (B.2)$$

so that it holds

$$\Gamma c = \oint dl \cdot m\dot{q} = 0 \qquad (B.3)$$

Moreover, since the action is contained in the exponential argument of the wave function, all the multiples of $2f\hbar$, with

$$S_{n(q,t)} = S_{0(q,t)} + 2nf\hbar = S_{0(q_0,t)} + \int_{q_0}^{q} dl \cdot p + 2nf\hbar \qquad n = 0, 1, 2, 3, ... \qquad (B.4)$$

are accepted.

### Quantum eigenstates in the hydrodynamic description

In the hydrodynamic description, the eigenstates are defined by the "equilibrium" condition

$$\dot{p} = 0 \tag{B.5.a}$$

that happens when the force generated by the quantum potential exactly counterbalances that one stemming from the Hamiltonian potential with the initial "stationary" condition

$$\dot{q} = 0. \tag{B.5.b}$$

The initial condition (B.5.b) united to the equilibrium condition leads to the stationarity $\dot{q} = 0$ along all times and, therefore, by (B.5.a) the eigenstates are irrotational.

Since the quantum potential changes itself with the state of the system, more than one stationary state (each one with its own $V_{qu_n}$) is possible and more than one quantized eigenvalues of the energy may exist.

For a time independent Hamiltonian $H = \frac{p^2}{2m} + V_{(q)}$, whose hydrodynamic energy reads [31] $E = \frac{p^2}{2m} + V_{(q)} + V_{qu}$, with eigenstates $\Psi_n(q)$ (for which it holds $p = m\dot{q} = 0$) it follows that

$$S_n = \int_{t_0}^{t} dt \left( \frac{p \cdot p}{2m} - V_{(q)} - V_{qu_n} \right) = -(V_{(q)} + V_{qu_n}) \int_{t_0}^{t} dt = -E_n(t - t_0) \tag{B.6}$$

where $V_{qu_n} = V_{qu}(\Psi_n)$, and that

$$V_{qu_n} = E_n - V_{(q)} \tag{B.7}$$

where (B.7) is the differential equation, that in the quantum hydrodynamic description, allows to derive to the eigenstates.

For instance, for a harmonic oscillator (i.e., $V_{(q)} = \frac{m\check{S}^2}{2}q^2$) (B.7) reads

$$V_{qu} = -(\frac{\hbar^2}{2m})/|\Psi_n|^{-1} \nabla \cdot \nabla |\Psi_n| = E_n - \frac{m\check{S}^2 q^2}{2}. \tag{B.8}$$

If for (B.8) we search a solution of type

$$|\Psi|_{(q,\,t)} = A_{n(q)} exp(-aq^2), \tag{B.9}$$

we obtain that $a = \frac{m\check{S}}{2\hbar}$ and $A_{n(q)} = H_n(\frac{m\check{S}}{2\hbar}q)$ (where $H_{n(x)}$ represents the $n$-th Hermite polynomial). Therefore, the generic $n$-th eigenstate reads

$$\Psi_{n(q)} = |\Psi|_{(q,\,t)} exp[\frac{i}{\hbar} S_{(q,t)}] = H_n(\frac{m\check{S}}{2\hbar}q) exp\left(-\frac{m\check{S}}{2\hbar}q^2\right) exp\left(-\frac{iE_n t}{\hbar}\right), \tag{B.10}$$

From (B.10) it follows that the quantum potential of the n-th eigenstate reads

$$V_{qu}{}^n = -(\frac{\hbar^2}{2m})|Œ/\nabla_q \bullet \nabla_q |Œ/$$

$$= -\frac{m\check{S}^2}{2}q^2 + \left[n\left(\frac{\frac{m\check{S}}{\hbar}H_{n-1} - 2(n-1)H_{n-2}}{H_n}\right) + \frac{1}{2}\right]\hbar\check{S} \qquad (B.12)$$

$$= -\frac{m\check{S}^2}{2}q^2 + (n+\frac{1}{2})\hbar\check{S}$$

where it has been used the recurrence formula of the Hermite polynomials

$$H_{n+1} = \frac{m\check{S}}{\hbar}qH_n - 2nH_{n-1}, \qquad (B.13)$$

that by (B.7) leads to

$$E_n = V_{qu_n} + V_{(q)} = (n+\frac{1}{2})\hbar\check{S}$$

The same result comes by the calculation of the eigenvalues that read

$$E_n = <Œ_n/H/Œ_n> = \int_{-\infty}^{\infty} Œ^*_{(q,t)} H^{op} Œ_{(q,t)} dq$$

$$= \int_{-\infty}^{\infty} |Œ|^2 \left[H_{(q,t)} + V_{qu}{}^n\right] dq$$

$$= \int_{-\infty}^{\infty} n_{(q,t)} \left[\frac{m}{2}\dot{q}^2 + \frac{m\check{S}^2}{2}(q-\underline{q})^2 + V_{qu}{}^n\right] dq \qquad (B.14)$$

$$= \int_{-\infty}^{\infty} n_{(q,t)} \left[\frac{1}{2m}\nabla S_{(q)}{}^2 + \frac{m\check{S}^2}{2}(q-\underline{q})^2 + V_{qu}{}^n\right] dq$$

$$= \int_{-\infty}^{\infty} n_{(q,t)} \left[\frac{m\check{S}^2}{2}(q-\underline{q})^2 - \frac{m\check{S}^2}{2}(q-\underline{q})^2 + (n+\frac{1}{2})\hbar\check{S}\right] dq = (n+\frac{1}{2})\hbar\check{S}$$

where $H^{op} = -\frac{\hbar^2}{2m}\frac{\partial^2}{\partial q^2} + V_{(q)}$ and where $n_{(q,t)} = Œ^*_{(q,t)}Œ_{(q,t)}$. Moreover, by applying (B.14) to (A.2-A.3) it follows that

$$\dot{p} = -\nabla(H + V_{qu}) = -\nabla((n+\frac{1}{2})\hbar\check{S}) = 0, \qquad (B.15)$$

$$\dot{q} = \frac{\nabla S_{(q,t)}}{m} = 0,  \qquad (B.16)$$

Confirming the stationary equilibrium condition of the eigenstates.

Finally, it must be noted that since all the quantum states are given by the generic linear superposition of the eigenstates (owing the irrotational momentum field $m\dot{q} = 0$) it follows that all quantum states are irrotational. Moreover, since the Schrödinger description is complete, do not exist others quantum irrotational states in the hydrodynamic description.

In the relativistic case, the hydrodynamic solutions are determined by the eigenstates $Œ^+{}_n, Œ^-{}_n$ derived by the irrotational stationary equilibrium condition applied to the momentum fields of matter and antimatter of equation (23), respectively.

# Appendix C

## The hydrodynamic HJE from the Lagrangian equation of motion

The identity

$$\frac{\partial L}{\partial \dot{q}^{\sim}} = p_{\sim} = \int_{t_0}^{t} \dot{p}_{\sim} dt = -\int_{t_0}^{t} \frac{\partial L}{\partial q^{\sim}} dt = -\frac{\partial}{\partial q^{\sim}} \int_{t_0}^{t} L dt = -\frac{\partial S}{\partial q^{\sim}} \qquad (C.1)$$

that stems from the equations (13-14), with the help of (10,12) leads to

$$p_{\sim} p^{\sim} = \frac{\partial S}{\partial q^{\sim}} \frac{\partial S}{\partial q_{\sim}} = \left(\frac{E^2}{c^2} - p^2\right)$$
$$= m^2 \chi^2 c^2 \left(1 - \frac{V_{qu}}{mc^2}\right) - m^2 \chi^2 \dot{q}^2 \left(1 - \frac{V_{qu}}{mc^2}\right) = m^2 c^2 \left(1 - \frac{V_{qu}}{mc^2}\right) \qquad (C.2)$$

that is the hydrodynamic HJE (1)

$$\frac{\partial S}{\partial q^{\sim}} \frac{\partial S}{\partial q_{\sim}} = m^2 c^2 \left(1 - \frac{\hbar^2}{m^2 c^2} \frac{\partial_{\sim} \partial^{\sim} /\!\!\!E/}{/\!\!\!E/}\right). \qquad (C.3)$$

# Appendix D

## The quantum potential in the region of space $R_0 < r \cong R_g$ with $R_0 \to R_g$

The balance between the quantum force and the gravitational one reads

$$\frac{du_{\sim}}{ds} = \frac{1}{2} \frac{\partial g_{\}\!\!\}\!\!\}}{\partial q^{\sim}} u^{\}} u^{|} - u_{\sim} \frac{d}{ds}\left(\ln\sqrt{1 - \frac{V_{qu}}{mc^2}}\right) + \frac{\partial}{\partial q^{\sim}}\left(\ln\sqrt{1 - \frac{V_{qu}}{mc^2}}\right) = 0 \qquad (D.1)$$

that by inserting the stationary condition (44) leads to

$$-\frac{1}{2}\frac{\partial g_{00}}{\partial q^1} = \frac{\partial}{\partial q^1}\left(\ln\sqrt{1 - \frac{V_{qu}}{mc^2}}\right) \qquad (D.2)$$

that in the vacuum space, for $r > R_0$, leads to

$$\frac{\partial}{\partial q^1}\left(\ln\sqrt{1 - \frac{V_{qu}}{mc^2}}\right) = -\frac{1}{2}\frac{\partial\left(1 - \frac{R_g}{r}\right)}{\partial q^1} \qquad (D.3)$$

and to

$$1 - \frac{V_{qu}}{mc^2} = exp\left[-\left(1 - \frac{R_g}{r}\right) + C_n\right] \qquad r > R_0 \qquad (D.4)$$

that gives

$$V_{qu} = mc^2\left(1 - exp\left[-\left(1 - \frac{R_g}{r}\right) + C_n\right]\right) \qquad r > R_0. \qquad (D.5)$$

Since $R_0 \leq R_g$ and since that for the minimum allowable mass we have that

$$R_0 \to R_g, \qquad (D.6)$$

for $R_0 < r <\approx R_g$, it follows that

$$mc^2\left(1 - exp[C_n]exp\left[-\left(1 - \frac{R_g}{R_0}\right)\right]\right) < V_{qu} \leq mc^2(1 - exp[C_n]) \qquad (D.7.a)$$

$$mc^2\left(1 - exp[C_n]\left[1 + \left(\frac{R_g - R_0}{R_0}\right)\right]\right) < V_{qu} \leq mc^2(1 - exp[C_n]) \qquad (D.7.b)$$

Moreover, since we are searching for the state with maximum mass concentration and hence with maximum quantum potential) from (D.7.b) it follows that this condition is achieved for $exp[C_n] = 0$ and, hence, for $C_n = -\infty$, that leads to

$$V_{qu} \cong mc^2.. \qquad (D.8)$$

Moreover, for $r = R_g + \vee$ with $\vee \ll R_g$ it follows that

$$\frac{mV_{qu}}{\hbar^2} = \frac{1}{/\!\!E/r^2}\partial^1 r^2\left(\left(\frac{R_g}{r} - 1\right)\partial_1 /\!\!E/\right) \cong \left(\frac{mc}{\hbar}\right)^2 \qquad (D.9)$$